\documentclass[twocolumn,epjc3]{svjour3}  

\RequirePackage{graphicx}
\usepackage{amssymb, amsmath}
\usepackage{upgreek}
\usepackage{color}
\newcommand{\Nuc}[2]{\ensuremath{^{#2}\mbox{#1}}}

\usepackage{tikz}
\newcommand{\len}{2.6} 
\newcommand{\scale}{1.75} 

\newcommand{\jpi}[2]{#1$^{#2}$}

\newcommand{\lev}[5]{
	\draw [level](#3,#1) -- (#3+\len, #1) ;
	\node[right] at (#3+\len +0.2,#1+#4) {#5, #2};
	}

\newcommand{\transition}[4]{
    \draw [->,color=gray, level](#1, #2) -- (#1,#3);
    \node [rotate=70, font=\tiny, fill=white, right ] at (#1,#2+0.04) {#4};
}

\usepackage{lineno}

\journalname{Eur. Phys. J. C}

\begin{document}

\title{Updated and novel limits on double beta decay and dark matter-induced processes in platinum}

\author{B.~Broerman\thanksref{e1,queens}
        \and
        M.~Laubenstein\thanksref{lngs}
        \and
        S.S.~Nagorny\thanksref{e2,queens,mi}
        \and
        S.~Nisi\thanksref{lngs}
        \and
        N.~Song\thanksref{e3,liverpool,itp}
        \and
        A.C.~Vincent\thanksref{queens,mi,pi}
}

\thankstext{e1}{E-mail: broerman@owl.phy.queensu.ca}
\thankstext{e2}{Email: sn65@queensu.ca}
\thankstext{e3}{Email: songnq@itp.ac.cn}

\institute{Department of Physics, Engineering Physics and Astronomy, Queen's University, Kingston, ON, K7L 3N6, Canada\label{queens}
           \and
           INFN -- Laboratori Nazionali del Gran Sasso, Assergi, I-67100, Italy\label{lngs}
           \and
           Arthur B. McDonald Canadian Astroparticle Physics Research Institute, Kingston, ON, K7L 3N6, Canada\label{mi}
           \and
           Department of Mathematical Sciences, University of Liverpool,  Liverpool, L69 7ZL, United Kingdom\label{liverpool}
           \and
           Institute of Theoretical Physics, Chinese Academy of Sciences, Beijing, 100190, China\label{itp}
           \and
           Perimeter Institute for Theoretical Physics, Waterloo, ON, N2L 2Y5, Canada\label{pi}
}

\date{Received: date / Accepted: date}

\maketitle

\begin{abstract}
A 510 day long-term measurement of a 45.3~g platinum foil acting as the sample and high voltage contact in an ultra-low-background high purity germanium detector was performed at Laboratori Nazionali del Gran Sasso (Italy). The data was used for a detailed study of double beta decay modes in natural platinum isotopes. Limits are set in the range $\mathcal{O}(10^{14} - 10^{19})$~yr (90\% C.L.) for several double beta decay transitions to excited states confirming, and partially extending existing limits. The highest sensitivity of the measurement, greater than $10^{19}$~yr, was achieved for the two neutrino and neutrinoless double beta decay modes of the isotope $^{198}$Pt. Additionally, novel limits for inelastic dark matter scattering on $^{195}$Pt are placed up to mass splittings of approximately 500~keV. We analyze several techniques to extend the sensitivity and propose a few approaches for future medium-scale experiments with platinum-group elements. 
\end{abstract}

\section{Introduction}\label{sec:intro}
Understanding the properties of the neutrino (e.g.~absolute mass scale, Dirac or Majorana particle nature, and scheme of the mass hierarchy) are among the goals of modern particle physics. One way to determine if the neutrino is its own antiparticle (Majorana in nature) is through the observation of neutrinoless double beta decay ($0\nu$-DBD). Unlike two neutrino double beta decay ($2\nu$-DBD) which is Standard Model process, $0\nu$-DBD is a lepton-violating process where the two simultaneously-generated neutrinos annihilate leaving only two outgoing beta particles. DBD-processes (double electron emission, $2\beta^-$, double positron emission, $2\beta^+$, double electron capture, $2\epsilon$, electron capture with positron emission, $\epsilon\beta^+$) can proceed to the ground state or to excited states of the daughter nucleus if energetically allowed. Current experimental efforts focus mostly on transitions to the ground state (see~\cite{dolinski2019neutrinoless} for a review). While the expected half-life of DBD-processes proceeding through transitions to the excited states are generally longer than to ground state transitions, some are predicted to be shorter due to resonant enhancement~\cite{bernabeu1983neutrinoless}. The possibility of DBD-processes with shorter half-lives is an attractive opportunity in the search for $0\nu$-DBD.

Current DBD-search experiments have sensitivity of $\mathcal{O}(10^{26})$~yr, aiming to probe the inverted mass hierarchy, using the technique where the DBD-active isotope is embedded in the sensitive detector volume. Some elements, especially within the platinum group (Ru, Pd, Os, Pt) are interesting for DBD studies but cannot be easily investigated due to the lack of high performance detector materials in which they can be embedded. The chemical properties of platinum-group elements make it difficult to attain a high mass fraction while satisfying the strict radiopurity requirements of low-background experiments. Therefore, these elements have been previously studied with an experimental approach ``source $\neq$ detector", where the metal sample is placed externally to a large volume low-background high-purity germanium (HPGe) detector. This method is limited to detecting the low-energy deexcitation $\gamma$'s that accompany DBD transitions to excited states of daughter nuclides~\cite{belli2009search,lehnert2011first,belli2021new,danevich2022new,danevich2020first,lehnert2016double}. However this approach has also met some experimental difficulties due to the reduced efficiency of $\gamma$ registration from non-optimized geometry of the experiment and self-absorption in the dense samples~\cite{danevich2022new,danevich2020first,lehnert2016double}.

Recently, a new technique replaces the traditional copper high voltage contact in a high purity germanium detector with a foil contact made of metal containing the isotope of interest~\cite{nagorny2021measurement,broerman2021search}. Here, the foil thickness ($\mathcal{O}(0.1)~$mm) and geometry can be optimized to maximize detection efficiency (i.e.~limit self-absorption).

Both \Nuc{Pt}{190} and \Nuc{Pt}{198}, which possess several DBD modes, have been investigated previously using Pt metal pieces of various thickness and geometries~\cite{belli2011first,danevich2022new}. While searches using this method were able to set limits on the half-life of $\mathcal{O}(10^{14} - 10^{19})$~yr, they suffer from low detection efficiency due to self-absorption of the low energy $\gamma$'s within the Pt sample itself. In the work presented here, data from a Pt foil sample using the new thin foil technique was analyzed to search for the various DBD modes. Comparable half-life limits are placed to~\cite{danevich2022new} of $\mathcal{O}(10^{14} - 10^{19})$~yr with less exposure due to the increase in detection efficiency. While these results, obtained with $\mathcal{O}(100)$-g-target mass are not competitive to those obtained in tonne-scale searches, it is useful to set limits with a diverse range of target isotopes, even through benchtop experiments, to aid in theoretical modeling as half-life predictions are obtained through model-dependent nuclear matrix element calculations.

The diversifying of target isotopes can also be helpful in the search for dark matter. Some well-motivated models of particle dark matter~\cite[and references therein]{harris2022snowmass} have effective couplings of the form $\chi_1 \chi_2 q \bar q$, where $\chi_1$ and $\chi_2$ need not have the same mass: one state can be seen as the excited state, and the other as the ground state. This leads to inelastic scattering, where the momentum transfer must overcome the mass splitting $\delta \equiv M_{\chi_2} – M_{\chi_1}$ between the two states in order for a recoil to occur. The lighter dark matter state $\chi_1$, being the dominant dark matter component, scatters inelastically with the nucleus and transitions to a heavier state $\chi_2$. Meanwhile, the nucleus is displaced and excited to a higher energy level. The excited nucleus subsequently deexcites to the ground state with the emission of a gamma, which can then be detected. The direct detection of inelastic dark matter (IDM) has been challenging, due to the kinematical suppression of the dark matter scattering rate in the detector arising from the mass splitting. At high mass splitting, the kinematics of the scattering imposes two conditions on the nucleus target: that the nucleus is heavy and the nuclear transition energy is low. 

As shown in~\cite{broerman2021search,song2021pushing,Bramante:2016rdh}, typical targets used in dedicated experiments (e.g. xenon, argon) are not heavy enough to allow for such an interaction for realistic halo dark matter velocities. Heavier nuclei such as Ta~\cite{lehnert2019search} and Pt offer the possibility of large momentum transfers and open the parameter space to larger mass splittings. In addition to recoil, dark matter scattering with Pt can yield a nuclear excitation—the absence of a deexcitation line thus yields a limit on the number of interactions with a heavy inelastic dark matter particle. Using \Nuc{Pt}{195}, with the relatively low first excited state at 98.9~keV, we probe inelastic dark matter up to the mass splitting of approximately 500~keV for a dark matter mass of 1 TeV. The search excludes the dark matter-nucleon cross section above approximately $10^{-33}$cm$^2$ to $10^{-28}$cm$^2$, pertinent to the mass splitting.

This paper is organized as follows: Section~\ref{sec:method} describes the experimental setup and sample radiopurity, Section~\ref{sec:bbsearch} the search for the DBD decay modes, Section~\ref{sec:dmsearch} a search for dark matter-induced deexcitations, and a discussion with conclusions in Section~\ref{sec:conc}.

\section{Experimental method}\label{sec:method}
\begin{figure}[h!]
\begin{center}
    \includegraphics[width=0.43\columnwidth]{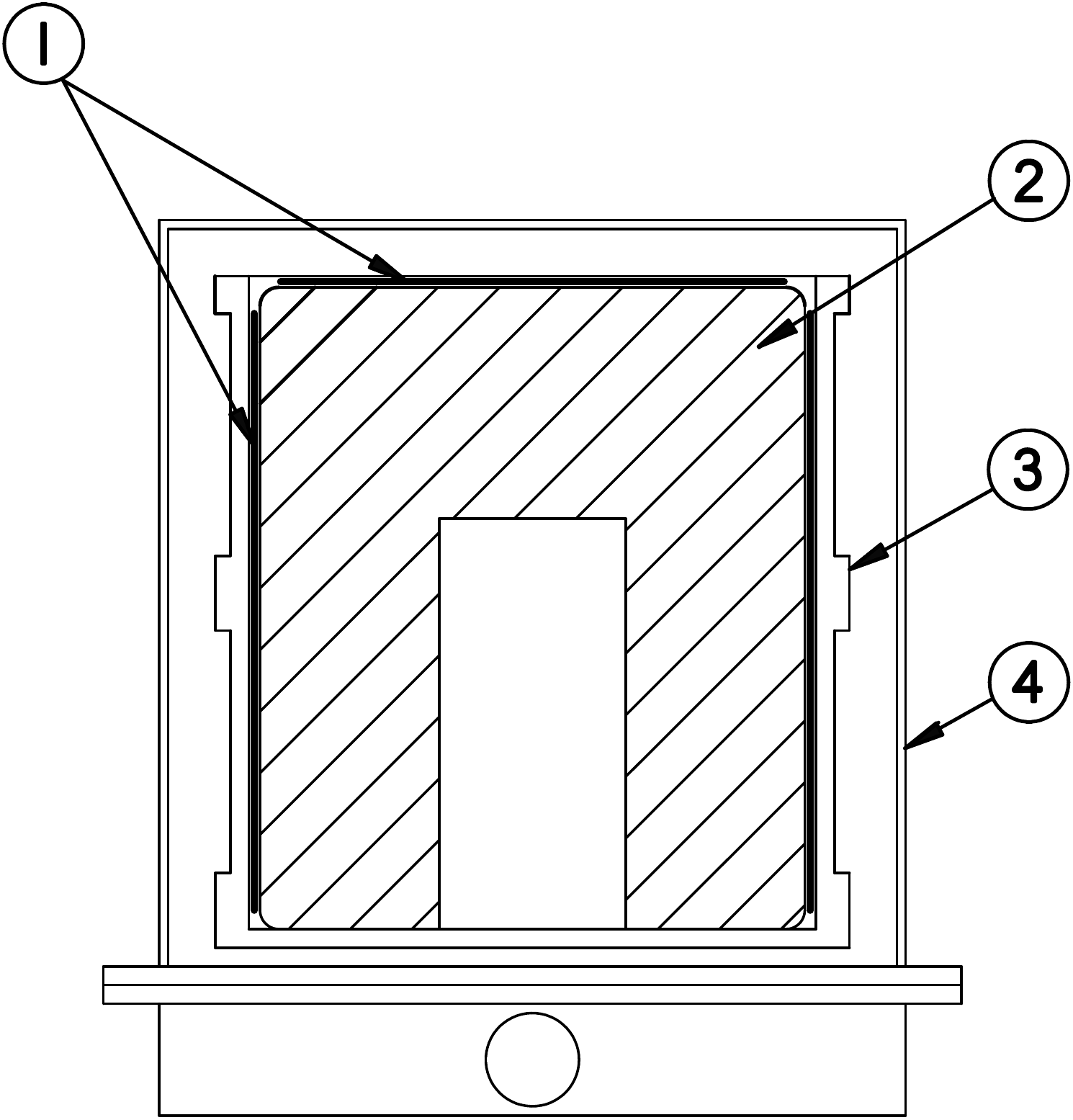}\includegraphics[width=0.57\columnwidth]{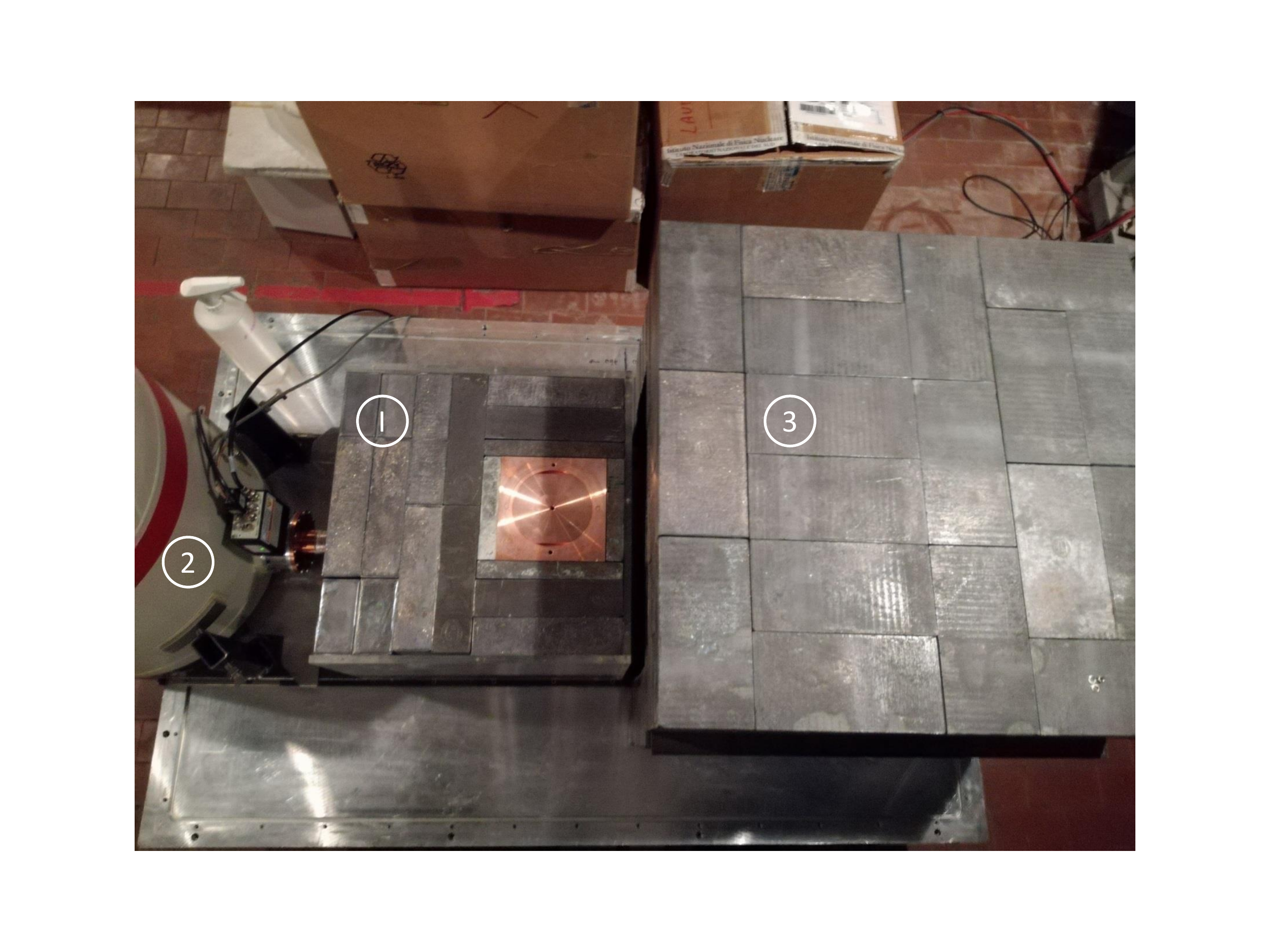}
\caption{Left: Section view of the HPGe detector and sample (not to scale) with 1) Pt foils on the top and wrapping the Ge crystal acting as the target and high-voltage contact, 2) HPGe crystal, 3) copper crystal holder, and 4) copper end cap of 1~mm thickness.
Right: Photo of the dedicated passive shield, in open configuration, shows 1) the movable part (dimensions ($40\times40\times40$)~cm) consisting of 5~cm of OFHC copper, 7~cm ULB lead, and up to 15~cm lead, all enclosed in a polymethylmethacrylate box continuously flushed with boil-off nitrogen,  2) the dewar fixed together with movable part on a stainless steel platform, 3) the fixed part of the passive shield with overall external dimensions ($60\times60\times80$)~cm.}
\label{fig:schematic}
\end{center}
\end{figure}

An ultra-low-background high-purity semi-coaxial p-type germanium detector (GS1) was operated underground at Laboratori Nazionali del Gran Sasso. A platinum foil of 0.12~mm thickness was wrapped around the 70-mm-diameter, 70-mm-high Ge detector serving as the sample and high voltage contact. An additional circular Pt foil, of a like diameter,  was placed on top of the Ge crystal. The sample had a mass of $(45.30 \pm 0.01)$~g and was made of commercially-available Pt foil of a 99.95\% purity, confirmed with a dedicated inductively-coupled plasma mass spectrometry measurement.  The detector and sample were placed inside the crystal holder made of oxygen-free high conductivity (OFHC) copper, and further placed in a dedicated passive multi-layer shield of 5~cm OFHC copper, 7~cm of ultra-low background lead, and 15 additional cm of low background lead. Details of the detector performance, data acquisition, and shielding setup are presented in~\cite{nagorny2021measurement}. A schematic of the sample, detector, and shielding are shown in Figure~\ref{fig:schematic}.

\begin{figure*}[h!]
\begin{center}
    \includegraphics[width=0.75\textwidth]{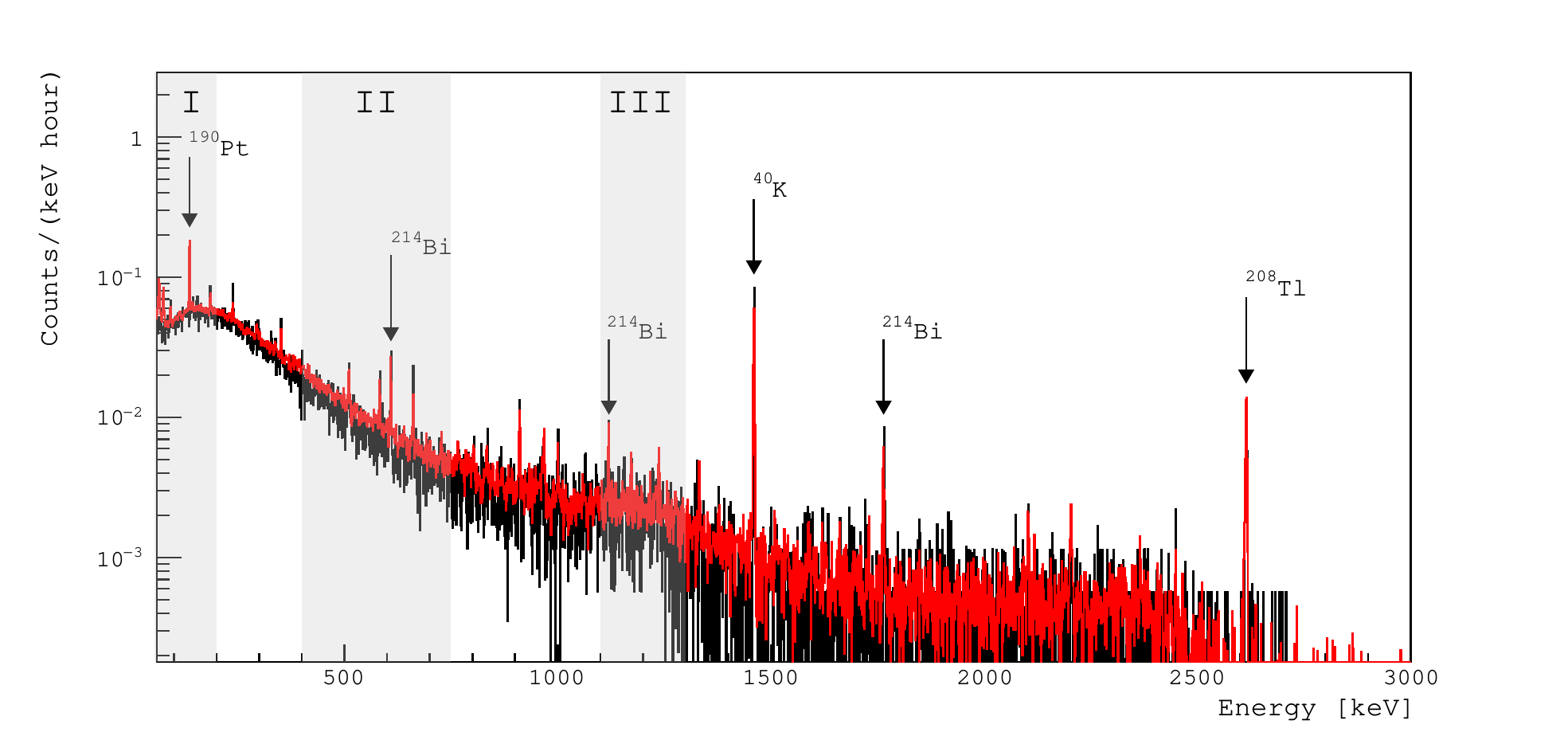}
\caption{Energy spectra from Pt foil sample data (red) and background data using a copper high voltage contact (black) normalized per hour of run time. Relevant leading background contaminants are indicated along with approximate energy regions of interest for the double beta decay search shaded. The specific decay mode energies in each region are given in Table~\ref{tab:bbresults}.}
\label{fig:fullSpectrum}
\end{center}
\end{figure*}

Approximately 1751~hr of background data using a copper high voltage contact and 12242~hr of data with the Pt foil were taken. The full energy spectra (counts/keV-hour) for the Pt foil sample and background data are shown in Figure~\ref{fig:fullSpectrum}. No cosmogenic nor anthropogenic radionuclides are observed in the Pt sample spectrum. Futhermore, no evidence of contamination from \Nuc{Ir}{192}, which was a major contribution to the backgrounds in~\cite{belli2011first} due to the chemically-similarity to Pt and half-life of approximately 73~days, is observed. In the Pt sample spectrum, gamma lines from natural radioactivity (\Nuc{U}{238}, \Nuc{U}{235}, and \Nuc{Th}{232} decay chains and \Nuc{K}{40}), \Nuc{Co}{60}, and \Nuc{Cs}{137} are observed. The peaks observed in the Pt sample spectrum, however, are consistent with background rates, leading to the upper limits (90\% C.L.) given in Table~\ref{tab:activity}.   

\begin{table}[h!]
\caption{Radiopurity of the Pt sample in mBq/kg, reproduced from~\cite{nagorny2021measurement}. Upper limits are given with 90\% C.L. The limit from \Nuc{Pb}{210} comes from a subset of data runs with a threshold sufficiently low be sensitive to the 46.5~keV gamma line. The activity of \Nuc{Ra}{226} is derived from \Nuc{Bi}{214} assuming secular equilibrium.}
\begin{center}
\begin{tabular}{lll}
Chain           & Isotope           & Activity [mBq/kg] \\ \hline
\Nuc{U}{238}    & \Nuc{Ra}{226}     & $< 0.81$ \\
                & \Nuc{Th}{234}     & $< 56$ \\
                & \Nuc{Pa}{234\rm{m}}& $< 31$ \\
                & \Nuc{Bi}{214}     & $< 0.81$ \\ \hline
\Nuc{U}{235}    & \Nuc{U}{235}      & $< 1.5$ \\ \hline
\Nuc{Th}{232}   & \Nuc{Ra}{228}     & $< 1.5$ \\
                & \Nuc{Th}{228}     & $< 1.0$ \\ \hline
                & \Nuc{Pb}{210}     & $< 26$ \\
                & \Nuc{Ir}{192}     & $< 0.51$ \\                 
                & \Nuc{Cs}{137}     & $< 0.31$ \\
                & \Nuc{Co}{60}      & $< 0.33$ \\
                & \Nuc{K}{40}       & $< 4.0$ \\ \hline
\end{tabular}\label{tab:activity}
\end{center}
\end{table}

\begin{table*}[h!]
\caption{Simplified decay scheme data of the \Nuc{Pt}{190} decay to \Nuc{Os}{190} and \Nuc{Pt}{198} decay to \Nuc{Hg}{198} for the energy levels investigated in this work. Isotopic abundance is taken from~\cite{meija2016isotopic}. The maximum available energy of transitions, Q$_{\beta\beta}$, spin/parity, and excited energy levels of the Os daughter are from~\cite{singh1990nuclear,xiaolong2009nuclear}.}
\begin{center}
\begin{tabular}{lllll}
Isotope         & Abundance & Q$_{\beta\beta}$ [keV] & Spin/parity & Excited energy \\ 
    &   &   &   &  level [keV] \\ \hline
\Nuc{Pt}{190}   & 0.012(1)  & 1401.3(4) & 0$^{+}$   & 1382.4    \\
                &           && 1, 2       & 1326.9    \\
                &           && 2$^{+}$   & 1114.7    \\
                &           && 0$^{+}$   & 911.8     \\
                &           && 2$^{+}$   & 558.0     \\
                &           && 2$^{+}$   & 186.7     \\
\Nuc{Pt}{198}   & 7.357(43) & 1050.3(21) & 2$^{+}$   & 411.8      \\
\end{tabular}\label{tab:levels}
\end{center}
\end{table*}

\section{Search for DBD modes}\label{sec:bbsearch}
In \Nuc{Pt}{190} we search for several DBD modes with different signatures. For $2\nu2\epsilon$ modes,
\begin{equation}
 e^- + e^- + (A,Z) \rightarrow (A, Z-2) + 2\nu + 2X\,,
\end{equation} 
\noindent the 2$\nu$'s escape the detector and one or both of the K/L shell X-rays can be detected. The mode is labeled depending on the K/L shell X-ray (e.g. if two K shell X-rays are emitted, the mode is labeled as $2\nu2$K). Due to the 50~keV energy threshold, we are sensitive to only X-rays from the K-shell; we look for only the K$_{\alpha1}$ emission at 63.0~keV. For the $0\nu2\epsilon$ modes,
\begin{equation}
e^- + e^- + (A,Z) \rightarrow (A, Z-2)^{*} \rightarrow (A,Z-2) + \gamma + 2X.
\end{equation}
\noindent Here, a bremsstrahlung photon or Auger electron is emitted along with two K or L shell X-rays. The energy of the bremsstrahlung photon is calculated as E$_{\rm{brem}}$ = Q$_{\beta\beta}$ - E$_{b1}$ - E$_{b2}$ - E$_{\gamma}$, for the maximum available energy of the transition Q$_{\beta\beta}$, and binding energies E$_{b1}$ (K = 73.871~keV) and E$_{b2}$ (L$_1$ = 12.968~keV) of the corresponding K/L shells of the daughter nuclide, and in the case of transition through the excited state, E$_\gamma$ is the deexcitation energy~\cite{barabash2020improved}. For the $0\nu2$K mode, E$_{\rm{brem}} = 1253.6$~keV, for $0\nu$KL E$_{\rm{brem}} = 1314.5$~keV, and E$_{\rm{brem}} = 1375.4$~keV for the $0\nu2$L mode.  Because of the lower background in the MeV-range with respect to the low energy X-ray region, we focus on the detection of the bremsstrahlung photon and do not consider coincidence between the bremsstrahlung photon and X-ray due to the low coincidence probability.

For $\epsilon\beta^+$ modes,
\begin{equation}
e^- + (A,Z) \rightarrow (A, Z-2) + e^+ + 2\nu(0\nu),
\end{equation} 
where 511~keV annihilation $\gamma$'s are produced from the emitted positron and can be detected. This signature is independent of the $2\nu/0\nu$ decay mode.

\begin{figure}
\begin{center}
\resizebox{7cm}{10cm}{
    \begin{tikzpicture}[
      scale=\scale, 
      level/.style={thick},
      virtual/.style={thick,densely dashed},
      trans/.style={thick,<->,shorten >=2pt,shorten <=2pt,>=stealth},
      classical/.style={thin,double,<->,shorten >=4pt,shorten <=4pt,>=stealth},
      label/.style = { font=\bf}
    ]
    \lev{0.00}{\jpi{(0)}{+}}{0}{0}{0.0};
    \lev{0.25}{(\jpi{2}{+})}{0}{0}{186.7};
    \lev{0.75}{(\jpi{4}{+})}{0}{0}{547.9};
    \lev{1.0}{(\jpi{2}{+})}{0}{0}{558.0};
    \lev{1.5}{(\jpi{3}{+})}{0}{0}{756.0};
    \lev{1.75}{(\jpi{0}{+})}{0}{0}{911.8};
    \lev{2.5}{(\jpi{2}{+})}{0}{0}{1114.7};
    \lev{3.0}{(\jpi{1,2}{})}{0}{0}{1326.9};
    \lev{3.25}{(\jpi{0}{+})}{0}{0}{1382.4};
    \transition{0.2}{3.25}{0.25}{1195.7 (100)}
    \transition{0.35}{3.}{0}{1326.9 (100)}
    \transition{0.5}{2.5}{0}{1114.7 (57)}
    \transition{0.65}{2.5}{0.25}{927.9 (100)}
    \transition{0.8}{2.5}{1.5}{358.7 (37)}
    \transition{0.95}{2.5}{1.75}{203.1 (6)}
    \transition{1.1}{1.75}{0.25}{725.1 (100)}
    \transition{1.25}{1.75}{1.0}{353.9 (28)}
    \transition{1.4}{1.5}{0.25}{569.3 (100)}
    \transition{1.55}{1.5}{0.75}{208.0 (4)}
    \transition{1.7}{1.5}{1.0}{197.9 (7)}
    \transition{1.85}{1.0}{0.}{558.0 (100)}
    \transition{2.0}{1.0}{0.25}{371.3 (73)}
    \transition{2.15}{0.75}{0.25}{361.1 (100)}
    \transition{2.3}{0.25}{0.0}{186.7 (100)}
    \node at (\len/2,-0.3) {${}^{190}_{76}$Os};

    \node at (3.55, 4.4) {${}^{190}_{78}$Pt};
    \draw (3.,4.6) -- (4, 4.6) ;
    \node at (4.5,4.6) {0.0, (\jpi{0}{+})};
    \draw [->](3.,4.6) -- (2.6, 4.1);
    \node at (3, 4.1) {$2\epsilon, \epsilon\beta^+$};
    \node at (3, 3.9) {Q$_{\beta\beta}$ = 1401.3(4)~keV};

    \draw[dashed] (0,-0.5) -- (4, -0.5);

    \draw (1.75,-1.5) -- (2.75,-1.5);
    \draw (1.75,-2.0) -- (2.75,-2.0);
    \node at (3.4, -1.5) {411.8, (\jpi{0}{+})};
    \node at (3.32, -2.0) {0.0, (\jpi{2}{+})};
    
    \transition{2.5}{-1.5}{-2.0}{411.8 (100)}    
    \node at (2.25,-2.3) {${}^{198}_{80}$Hg};

    \node at (0.75, -1.0) {${}^{198}_{78}$Pt};
    \draw (0.25,-0.75) -- (1.25, -0.75);
    \node at (1.75,-0.75) {0.0, (\jpi{0}{+})};
    \draw [->](1.25,-0.75) -- (1.65, -1.25);   
    \node at (0.75, -1.4) {Q$_{\beta\beta}$ = 1050.3(21)~keV};

    \end{tikzpicture}
    }
    \caption{Energy level diagrams for \Nuc{Pt}{190} decay to \Nuc{Os}{190} (top) and \Nuc{Pt}{198} decay to \Nuc{Hg}{198} (bottom).} \label{fig:Elevels}
   \end{center}
\end{figure}

The $\gamma$/X-ray energies, $E_{\gamma,X}$, searched for in this work can be bound into three regions of interest, shaded in Figure~\ref{fig:fullSpectrum}. Since all double beta decay isotopes have a $0^+$ ground state, the most likely transitions are to a $0^+$ states of the daughter. Spin suppression occurs for larger angular momentum transfers. Hence, we limit our search for double beta decays to the first excited $0^+$ state and low-lying $2^+$ states. Energy level diagrams for the DBD of \Nuc{Pt}{190} and \Nuc{Pt}{198} are shown in Figure~\ref{fig:Elevels}.

\begin{figure}[h!]
\begin{center}
    \includegraphics[width=\columnwidth]{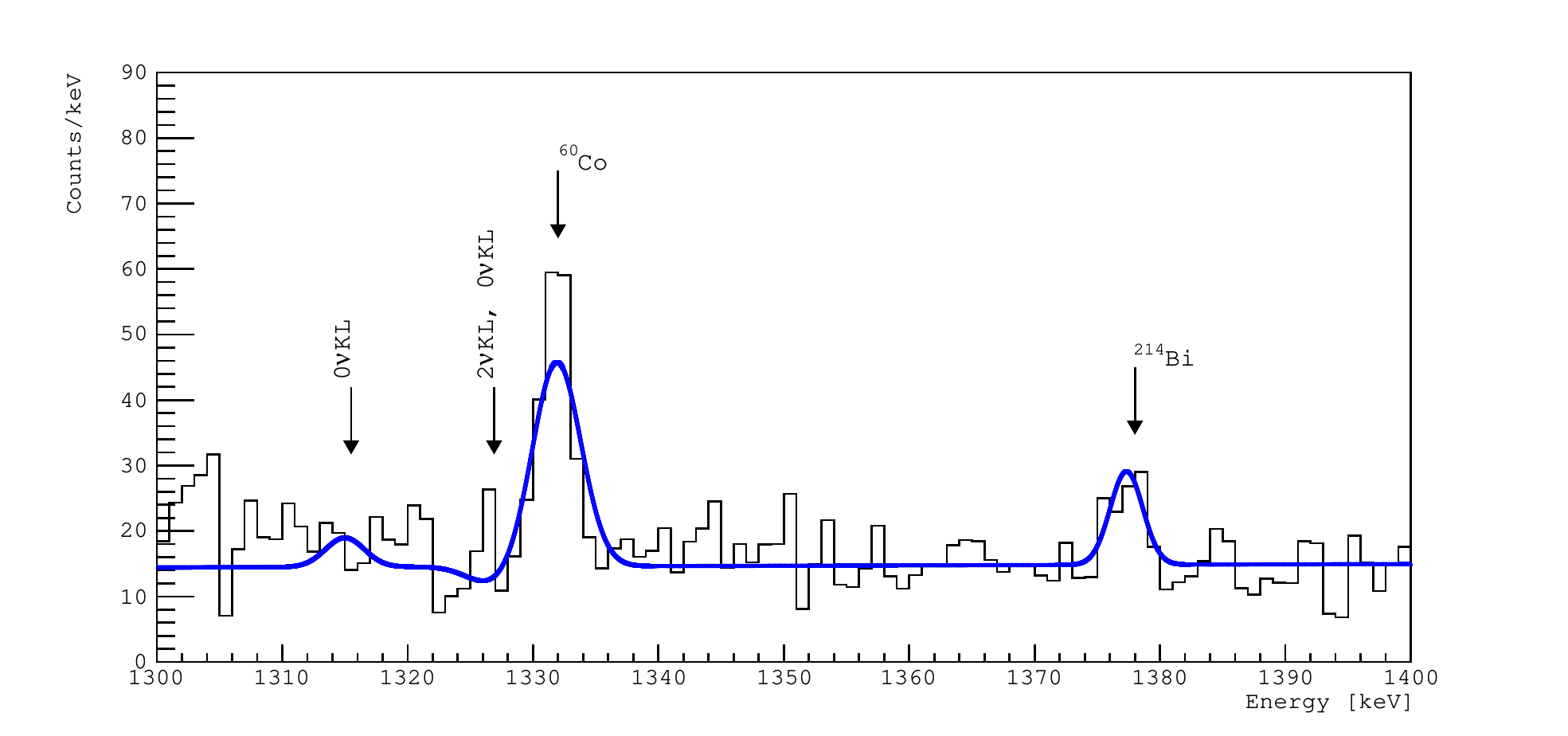}
\caption{Example fit to region \textsc{III} in Figure~\ref{fig:fullSpectrum} sensitive to the $2\nu$KL excited state and resonant $0\nu$KL excited state modes at 1326.9~keV and $0\nu$KL ground state mode from $1314.5$~keV of \Nuc{Pt}{190}. Background peaks are observed for \Nuc{Co}{60} at 1332~keV and \Nuc{Bi}{214} at 1378~keV.}
\label{fig:spectrumFit}
\end{center}
\end{figure}

The Pt sample energy spectrum was fitted in each region of interest with a model describing the background (linear in the case of featureless and linear + gaussian(s) in the presence of known background gamma lines) and effect (gaussian). An example fit is shown in Figure~\ref{fig:spectrumFit}. Using the fit results and following the Feldman-Cousins method~\cite{feldman1998unified}, we obtain a limit on the number of counts, lim S, and calculate the corresponding limit on $T_{1/2}$ at 90\% C.L.

No evidence of DBD decays to any studied excited states is observed. An upper limit on the half-life, $\rm{T}_{1/2}$ for each DBD mode searched for is calculated as, 

\begin{equation}
\rm{lim}~T_{1/2} = \ln(2) \cdot \mathcal{N} \cdot \eta \cdot t/\rm{lim}~S,
\end{equation}
where $\mathcal{N}$ is the number of beta decaying atoms in the sample, $\eta$ is the detection efficiency for the energy of interest, and $t$ is the counting time. We assume natural isotopic composition of platinum given in Table~\ref{tab:levels}. The detection efficiency is calculated with a GEANT4~\cite{agostinelli2003geant4} simulation and includes the decay scheme of the daughter nuclide. 

The limits for $\rm{T}_{1/2}$ calculated in this analysis are shown in Table~\ref{tab:bbresults} and are strongly dependent on the isotopic abundance and the detection efficiency. In general, the results presented here are comparable to leading limits from~\cite{danevich2022new}, albeit with less sample mass. The highest sensitivity we achieved is of \Nuc{Pt}{198} at $\geq1.5\times10^{19}$~yr due to the larger isotopic abundance of \Nuc{Pt}{198} over \Nuc{Pt}{190}. Indeed, the $\sim 500\times$ increase in half-life sensitivity for the \Nuc{Pt}{198} decay over \Nuc{Pt}{190} decay modes with positron emission (e.g. $2\nu\epsilon\beta^+$) with a similar region of interest closely matches the $\sim500\times$ abundance of \Nuc{Pt}{198}  over \Nuc{Pt}{190}. However, despite our optimized sample geometry and high detection efficiency, we are limited by the internal background of the detector setup which result in limits which are not consistently stronger than derived in~\cite{danevich2022new}. In particular, we set stronger limits for $0\nu2$K, $0\nu$KL, and $0\nu2$L to the ground state, $0\nu$KL resonant to the excited state, and $2\nu$KL to the excited state.

\begin{table*}[h!]\label{tab:bbresults}
\caption{Limits on the DBD half-lives for \Nuc{Pt}{190} and \Nuc{Pt}{198}. Given are the transitions, decay modes to ground state (g.s.) or excited state (exc.), and corresponding $\gamma/$X-ray energy, detection efficiency, and experimental $\rm{T}_{1/2}$ limits (90\% C.L) derived here and compared to~\cite{danevich2022new}.}
\begin{center}
\begin{tabular}{llllll}
        &        &                       &               & \multicolumn{2}{c}{Experimental $\rm{T}_{1/2}$ [$\times 10^{16}$~yr]}  \\
Transition & Decay Mode      & $E_{\gamma,X}$ [keV]    & $\eta$      & This work     & \cite{danevich2022new} \\ \hline
                                            &&&&&\\
\Nuc{Pt}{190} $\rightarrow$ \Nuc{Os}{190}&$2\nu$2K (g.s.) & 63.0          & 0.175         & $\geq 0.08$   & $\geq 0.10$ \\
&$2\nu$KL (g.s.) & 63.0          & 0.175         & $\geq 0.08$   & $\geq 0.052$ \\
&$0\nu2$K (g.s.) & 1253.6      & 0.18	        & $\geq 7.7$    & $\geq 4.7$ \\
&$0\nu$KL (g.s.) & 1314.5     & 0.18          & $\geq 16.5$   & $\geq 4.6$ \\
&$0\nu2$L (g.s.) & 1375.4     & 0.18          & $\geq 3.8$    & $\geq 3.2$ \\ \\
&$2\nu$KL (exc.) & 1326.9                & 0.18	        & $\geq 15.9$   & $\geq 2.5$ \\
&Res. $0\nu$KL (exc.)    & 1326.9        & 0.18	        & $\geq 15.9$   & $\geq 2.5$ \\
&Res. $0\nu$KL (exc.)    & 1195.7        & 0.18          & $\geq	9.3$    & $\geq 4.9$ \\ \\
&$2\nu2\epsilon$ (exc.) & 186.7          & 0.093         & $\geq 1.2$    & $\geq 0.98$ \\
&$2\nu2\epsilon$ (exc.) & 558.0          & 0.087         & $\geq 2.4$    & $\geq 2.3$ \\
&$2\nu2\epsilon$ (exc.) & 725.1          & 0.072         & $\geq	1.0$    & $\geq 5.2$ \\
&$2\nu2\epsilon$ (exc.) & 1114.7         & 0.047         & $\geq 1.5$    & $\geq 1.9$ \\
&$0\nu2\epsilon$ (exc.)  & 186.7         & 0.05          & $\geq 0.69$   & $\geq 0.76$ \\
&$0\nu2\epsilon$ (exc.)  & 558.0         & 0.05          & $\geq 1.4$    & $\geq 1.9$ \\
&$0\nu2\epsilon$ (exc.)  & 725.1         & 0.04          & $\geq 0.58$   & $\geq 4.8$ \\
&$0\nu2\epsilon$ (exc.)  & 1114.7        & 0.034         & $\geq 1.1$    & $\geq 1.6$ \\ \\
&$2\nu\epsilon\beta^+$ (g.s.) & 511      & 0.32          & $\geq	3.2$    & $\geq 2.9$ \\
&$0\nu\epsilon\beta^+$ (g.s.) & 511      & 0.31	        & $\geq 3.0$    & $\geq 2.9$ \\
&$2\nu\epsilon\beta^+$ (exc.) & 511      & 0.22          & $\geq 2.2$    & $\geq 2.8$ \\
&$0\nu\epsilon\beta^+$ (exc.) & 511      & 0.22  	    & $\geq 2.2$    & $\geq 2.8$ \\
                                            &&&&&\\
\Nuc{Pt}{198} $\rightarrow$ \Nuc{Hg}{198}&$2\beta^- (2\nu+0\nu)$ (exc.) & 411.8   & 0.24          & $\geq 1500$   & $\geq 3200$ \\ \hline
\end{tabular}
\end{center}
\end{table*}

\section{Inelastic dark matter search}\label{sec:dmsearch}
We focus on the most abundant isotope of natural platinum, \Nuc{Pt}{195}, to search for inelastic dark matter scattering from nuclear excitation. The scattering rate, $\mathcal{R}$, is computed as~\cite{song2021pushing}

\begin{equation}
    \mathcal{R}= N_{T}\dfrac{\rho_\chi}{M_\chi}\int_{v_{\min}}^{v_{\max}} dv vf(v) \int_{E_{R,{\min}}}^{E_{R,{\max}}} \dfrac{d\sigma_{\chi N}}{dE_R} dE_R\,,
    \label{eq:eventrate}
\end{equation}
\noindent where

\begin{equation}
    \dfrac{d\sigma_{\chi N}}{dE_R}=\dfrac{\sigma_n M_N}{2v^2\mu_{\chi n}^2}S(q)\,,
    \label{eq:dsigmachiN}
\end{equation}
\noindent with $N_T$ the number of target nuclei, $M_\chi$ ($M_N$) the mass of the dark matter particle (nucleus), and $\mu_{\chi n}$ the reduced mass between the dark matter particle and the nucleon. The minimum and the maximum recoil energy, $E_{R,{\min}}$ and $E_{R,{\max}}$, respectively, dependent on the velocity $v$ ranging from $v_{\min}$ and $v_{\max}$, is related to the nuclear excitation energy and the mass splitting, as seen from~\cite{song2021pushing}. We have assumed a Maxwellian dark matter distribution, $f(v)$, with $v_0=220$~km/s, the Earth velocity $v_e=240$~km/s, and the local escape velocity at the position of the Earth $v_{esc}=600$~km/s~\cite{Monari:2018ckf}. The local dark matter energy density is assumed to be $\rho_\chi=0.4$~GeV/cm$^3$, consistent with the recent determination of the local dark matter density~\cite{deSalas:2020hbh}. The dark matter nuclear scattering cross section is related to the per nucleon scattering cross section $\sigma_n$ by Equation~\eqref{eq:dsigmachiN}. In the absence of coherent enhancement, the nuclear response $S(q)$ is estimated to be
\begin{equation}
    S(q) = \dfrac{A^2}{Z^2}(2J_i+1)j_2(qR)^2\dfrac{B(E2)}{e^2R^4}\,,
    \label{eq:SE2}    
\end{equation}
\noindent for atomic number $A$, proton number $Z$, initial nuclear spin $J_i$, and momentum-dependent (momentum transfer $q$ and atomic radius $R$) Bessel function of the second kind $j_2(qR)$. For \Nuc{Pt}{195} the value of the reduced transition probability $B(E2)$ is measured to be 11.1 W.u.~\cite{huang2014nuclear} and the initial nuclear spin $J_i=1/2$. 

\begin{figure}[h!]
\begin{center}
    \includegraphics[width=\columnwidth]{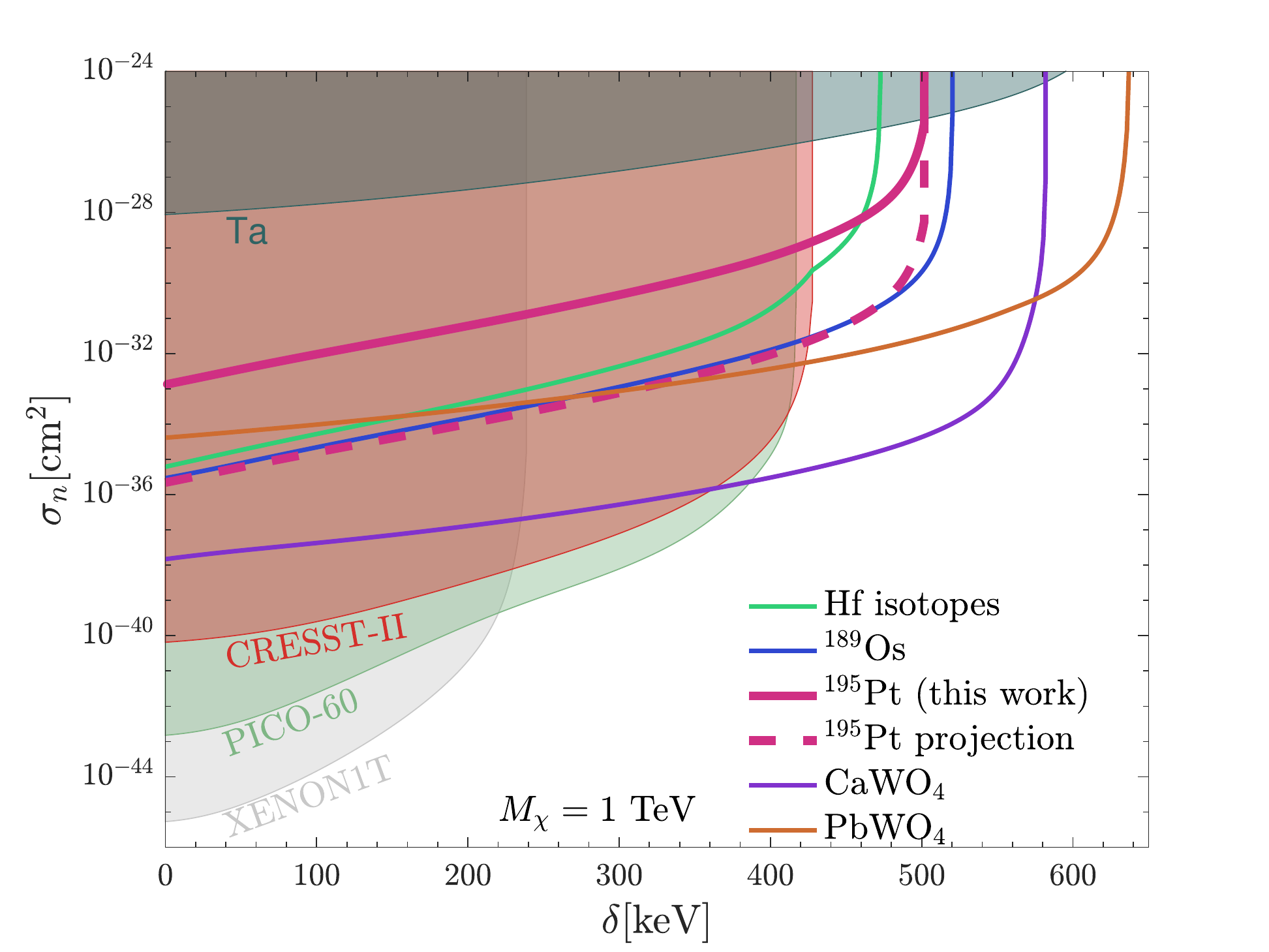}
\caption{Constraints on inelastic dark matter-nucleon scattering cross section at $90\%$ C.L. assuming a dark matter mass $M_\chi=1$~TeV. We also show limits based on data from PICO-60~\cite{Amole:2015pla}, CRESST-II~\cite{Angloher:2015ewa}, XENON1T~\cite{Aprile:2018dbl}, \Nuc{Ta}{180}~\cite{lehnert2019search}, Os~\cite{Belli:2020vnc}, CaWO$_4$~\cite{Munster:2014mga} and PbWO$_4$~\cite{Beeman:2012wz}, adapted from~\cite{song2021pushing} and the Hf constraint from~\cite{broerman2021search}. The solid magenta line depicts the limit on cross section from the detection of the deexcitation gamma when inelastic dark matter scatters and excites the \Nuc{Pt}{195} nucleus, derived in this work. The dashed magenta line shows the projected limit assuming cumulative increases in sensitivity described in Section~\ref{sec:conc}.}
\label{fig:Ptlimit}
\end{center}
\end{figure}

The measured event spectrum is quite flat around the excitation energy 98.9 keV with a rate $s=654$~keV$^{-1}$~\cite{nagorny2021measurement}. Therefore, the dark matter contribution should not exceed the variance of the measured background rate. At 68\% C.L., this is
\begin{equation}
    \mathcal{R}_\mathrm{Bkg}=\sqrt{s\cdot\sigma_\epsilon}/(\eta \cdot t).
\end{equation}
\noindent Here, with the energy resolution $\sigma_\epsilon=1.90\pm 0.25$~keV, detection efficiency $\eta=5.5\%$, and the counting time $t=12242$~hr we obtain $\mathcal{R}_\mathrm{Bkg}=1.45\times 10^{-5}$ keV$^{-1}$~s$^{-1}$. The limit on inelastic dark matter-nucleon scattering is shown as a function of the mass splitting $\delta$ in Figure~\ref{fig:Ptlimit} by requiring the dark matter scattering rate not to exceed the measured background rate, i.e., $\mathcal{R}\leq 1.645\mathcal{R}_\mathrm{Bkg}$, with the numerical factor accounting for  the conversion from 68\% to 90\% confidence level. As demonstrated in~\cite{song2021pushing}, a larger mass splitting indicates smaller range of kinematically allowed recoil energy, and hence smaller dark matter scattering rate from Eq.~\eqref{eq:eventrate}, which weakens the bounds on cross section. The result extends the previous direct detection constraints derived from PICO-60~\cite{Amole:2015pla}, CRESST-II~\cite{Angloher:2015ewa}, XENON1T~\cite{Aprile:2018dbl} for the mass splitting from 430~keV to 500~keV. The new limit, however, remains less stringent to constraints based on \Nuc{Os}{189}~\cite{Belli:2020vnc}, CaWO$_4$~\cite{Munster:2014mga}, and PbWO$_4$~\cite{Beeman:2012wz} derived in~\cite{song2021pushing}.

\section{Discussion and conclusions}\label{sec:conc}
A search for rare nuclear processes that can occur in natural platinum isotopes was performed using an ultra-low-background HPGe detector and a 45.3~g Pt metal foil sample. The search included double beta decays, double electron captures into excited states of the daughter isotopes, and dark matter induced events. No signal was found and 90\% confidence level limits on the half-life of the different DBD processes of $\mathcal{O}(10^{14}-10^{19})$~yr were set. Existing constraints on double beta decay and double electron capture modes were confirmed and partially improved. We additionally place novel inelastic dark matter limits searching for the deexcitation gamma from the 98.9~keV excited state of \Nuc{Pt}{195} up to mass splitting of approximately 500 keV excluding the dark matter-nucleon cross section above the range $10^{-33}-10^{-28}$~cm$^2$ for a dark matter mass of 1~TeV. While not as competitive as existing limits, these results diversify the list of target materials used in dark matter searches.  

General improvement of this measurement can be achieved through further background reduction and increased sample mass and measurement time. Indeed, the content of radionuclides of the market-available platinum metal was already very low with no significant contribution from the natural decay chains \Nuc{Th}{232}, \Nuc{U}{235}, and \Nuc{U}{238}, nor from \Nuc{K}{40}, \Nuc{Co}{60}, and \Nuc{Cs}{137}. 

The total acquisition time presented here of approximately 510~days could be significantly improved in future measurements. A 5~year measurement would increase the sensitivity by approximately a factor of 2. 

To increase the mass of the sample there are several strategies that could be realized within the ``source $\neq$ detector" approach. For example, a promising method is to use a larger Pt sample mass with a HPGe detector array as adopted by the TGV collaboration~\cite{rukhadze2011new}. There, a thin foil of the studied material is placed between 16 pairs of Ge detectors (20.4~cm$^2 \times 0.6$~cm) stacked in a large tower to search for different modes of $2\epsilon$ processes in \Nuc{Cd}{106}. Assuming the same area (20.4~cm$^2$) and a Pt foil thickness of 0.12~mm, as used in our study, the total Pt sample mass would be 84~g. This would further increase the sensitivity by approximately a factor of 2. Moreover, by exploiting the coincidence between neighboring, face-to-face detectors, this method can strongly suppress backgrounds (by a factor of 10) and increase detection efficiency (by a factor of 2) leading to an even further increase in sensitivity. Unfortunately, HPGe-based experimental methods have a limitation coming from the finite energy resolution (FWHM~$\approx 2.5$~keV in the $20-100$~keV range for both~\cite{rukhadze2011new} and our study).

One of the possible further modification of the stacked detector array approach would be to use tower of Ge wafers working as cryogenic detectors, similar to the light detectors utilized in the CUPID-0~\cite{beeman2013characterization} or CUPID-Mo~\cite{novati2019charge} experiments. Here, the Pt metal foil samples would be placed between neighboring, face-to-face wafers. Typical energy resolution for cryogenic light detectors are $\mathcal{O}(300)$~eV FWHM~\cite{beeman2013characterization} that can be further improved to $\mathcal{O}(50)$~eV using Neganov-Trofimov-Luke amplification~\cite{novati2019charge}. The improved energy resolution would help to minimize background contributions to the region of interest leading to a factor of $10-50$ experimental sensitivity enhancement. This method may also be useful in future inelastic dark matter searches with mass splitting above 400~keV due to a wide variety of target isotopes that can be studied. Recently, a prototype based on a similar stacked-wafer approach has been developed and tested using eight large-area ($\varnothing$ 150~mm) Si wafers to study surface alpha contamination~\cite{benato2023development}.

However, to approach the half-life sensitivity of $\mathcal{O}(10^{24})$~yr, a drastic improvement of detection efficiencies is necessary. Realistically, this can be achieved in several ways though the ``source = detector” approach, where target isotopes would be embedded into the detector material. Recently, scintillating crystals of the Cs$_2$XCl$_6$ family started to be extensively studied because of their high light yield and good energy resolution, as well as due to their low internal background, see~\cite{nagorny2021novel}. Compounds of this family are extremely flexible to accommodate different elements (X = Ru, Pd, Os, Pt). This approach would especially benefit searches for low energy $\gamma$'s where multiple orders of magnitude improvement in detection efficiency can be achieved.

A well-known method to significantly increase the number of decaying nuclei is to use materials enriched with the isotope of interest, as performed in the LEGEND~\cite{abgrall2017large}, CUPID-0~\cite{armatol2022toward}, and CUPID-Mo~\cite{augier2022final} experiments. However, this becomes difficult for elements of the Pt group as they can only be enriched by electromagnetic separation, a very expensive and time-consuming process. Price-evaluation performed on elements of this group with similar initial isotopic abundance of $\mathcal{O}(0.01)$\% gives a value of \$1000/mg for a final enrichment level of $\mathcal{O}(10)$\%. With adequate funding, one would be able to enhance sensitivity to a factor of 1000. 

Despite initial complications in the study of double beta decay processes in natural isotopes of elements in platinum group, a number of techniques now exist to perform these investigations. With adaptation and tuning of the methods described above, even further sensitivity is possible. 

\begin{acknowledgements}
B.B.~is supported by the Natural Sciences and Engineering Research Council of Canada (NSERC). N.S.~would like to thank the UK Science and Technology Facilities Council (STFC) for funding this work through support for the Quantum Sensors for the Hidden Sector (QSHS) collaboration under grants ST/T006102/1, ST/T006242/1, ST/T006145/1, ST/T006277/1, ST/T006625/1, ST/T006811/1, ST/T006102/1 and ST/T006099/1. S.N. and A.C.V.~are supported by NSERC and the Arthur B. McDonald Canadian Astroparticle Physics Research Institute.
\end{acknowledgements}

\bibliographystyle{spphys} 
\bibliography{02_bibliography}

\end{document}